\documentclass[num-refs]{wiley-article}

\usepackage{graphicx}
\usepackage[space]{grffile}
\usepackage{latexsym}
\usepackage{textcomp}
\usepackage{longtable}
\usepackage{tabulary}
\usepackage{booktabs,array,multirow}
\usepackage{amsfonts,amsmath,amssymb}
\usepackage{natbib}
\usepackage{url}
\usepackage{hyperref}
\hypersetup{colorlinks=false,pdfborder={0 0 0}}
\usepackage{etoolbox}
\makeatletter
\patchcmd\@combinedblfloats{\box\@outputbox}{\unvbox\@outputbox}{}{%
  \errmessage{\noexpand\@combinedblfloats could not be patched}%
}%
\makeatother

\usepackage[utf8]{inputenc}
\usepackage[english]{babel}

\corraddress{Ma\l gorzata Olejniczak, Centre of New Technologies, University of Warsaw, S. Banacha 2c, 02-097 Warsaw, Poland}
\corremail{m.olejniczak@cent.uw.edu.pl}
\fundinginfo{
Polish National Science Centre (2016/23/D/ST4/03217), 
Labex CaPPA (ANR-11-LABX-0005-01) and I-SITE ULNE project OVERSEE (ANR-16-IDEX-0004), 
CPER CLIMIBIO (European Regional Development Fund, Hauts de France council, French  Ministry of Higher
Education and Research), the French national supercomputing facilities (DARI A0050801859), the European Commission grant H2020-FETHPC-2017 ``VESTEC'' (ref. 800904)
}

\papertype{Full Paper}

\title{A Topological Data Analysis Perspective on Non-Covalent Interactions in Relativistic Calculations}

\author[1]{Ma\l gorzata Olejniczak}
\author[2]{Andr\'e Severo Pereira Gomes}
\author[3]{Julien Tierny}

\affil[1]{Centre of New Technologies, University of Warsaw, S. Banacha 2c, 02-097 Warsaw, Poland}
\affil[2]{Universit\'e de Lille, CNRS, UMR 8523 -- PhLAM -- Physique des Lasers, Atomes et Mol\'ecules, F-59000 Lille, France}
\affil[3]{Sorbonne Universit\'e, CNRS, Laboratoire d'Informatique de Paris 6, LIP6, F-75005 Paris, France}

\begin{document}

\maketitle

\begin{abstract}
Topological Data Analysis (TDA) is a powerful mathematical theory, largely unexplored in theoretical chemistry. In this work we demonstrate how TDA provides new insights into topological features of electron densities and reduced density gradients, by investigating the effects of relativity on the bonding of the Au$_4$-S-C$_6$H$_4$-S'-Au'$_4$ molecule. Whereas recent analyses of this species carried out with the Quantum Theory of Atoms-In-Molecules [Anderson \textit{et al.}, Chem.\ Eur.\ J.\ 25, 2538, 2019] concluded, from the emergence of new topological features in the electron density, that relativistic effects yielded non-covalent interactions between gold and hydrogen atoms, we show from their low persistence values (which decrease with increased basis set size) these features are not significant. Further analysis of the reduced density gradient confirms no relativity-induced non-covalent interactions in Au$_4$-S-C$_6$H$_4$-S'-Au'$_4$. We argue TDA should be integrated into electronic structure analysis methods, and be considered as a basis for the development of new topology-based approaches.

\end{abstract}

\section{Introduction}
A well-exploited consequence of the Hohenberg-Kohn theorem\cite{hohenberg-pr-136-b864-1964} is that the exact, non-degenerate, ground-state electron density (ED) contains complete molecular information and therefore encodes all interactions in molecular systems. 
In order to retrieve information on interatomic interactions from the ED, this property is subjected to various analysis schemes, which often involve visual exploration. This is, however, both a technical challenge and a test of chemical concepts. 
The former relates to a nontrivial choice of analysis techniques of scientific data. For instance, \emph{direct} visualization schemes (an example of which is representing 3D scalar fields as isosurfaces),
although intuitive and simple in realization, often lead to images that are cluttered and difficult to grasp, especially when aimed to illustrate complex scientific questions. On the other hand, the so-called \emph{feature-based} visualization methods are designed to recapture only the meaningful features of data, however are more complex to comprehend and therefore not as widely used.\cite{lipsa-cgf-31-2317-2012,tong-information-9-65-2018}

Topological Data Analysis is a recent research field, at the interface 
between mathematics and computer science, which suggests to investigate data 
based on its structure.\cite{heine16}
Rooted in sound theoretical 
settings such as Morse theory,\cite{milnor63} 
TDA provides to users a set of tools for the robust extraction 
of topological features in their data. This includes for instance  
critical points, integral lines, separating surfaces, voids, etc. Moreover, 
Persistent Homology (PH) \cite{edelsbrunner09} brings an appealing framework for 
measuring the salience of topological features in the data, which is well 
established both at a theoretical and practical level.\cite{otter-epjds-2017} In practice, it allows 
users to discriminate important features from non-significant configurations 
and provides the necessary basis for multi-scale data analysis.\cite{tierny17} Recent years have witnessed some successful applications of TDA to chemistry, where it helped to understand hydrogen-bonding networks in ion and molecule aggregates,\cite{xia-pccp-20-13448-2018,xia-pccp-21-21038-2019,steinberg-jc-11-48-2019} aqueous solubility of molecules,\cite{wu-jcc-39-1444-2018,pirashvili-jc-10-54-2018} stability of fullerenes,\cite{xia-jcc-36-408-2015} molecular transition pathways,\cite{beketayev-cgf-30-663-2011} conformational spaces of molecules\cite{membrillo-solis-arxiv-2019} or the bonding patterns in molecular systems.\cite{weber-ieeetvcg-13-330-2007,gunther-ieeetvcg-20-2476-2014,hermosilla-ieeetvcg-23-731-2017,bhatia-jcc-39-936-2018}

In this manuscript we focus on the use of TDA to investigate non-covalent interactions (NCIs) for systems containing heavy elements. The description of NCIs is by itself a challenge for existing chemistry concepts, and has been a driving force for the development of new analysis techniques.\cite{pastorczak-jcp-146-120901-2017,clark-pccp-20-30076-2018} For systems containing heavy elements, there is the additional challenge of 
 including the effects of relativity in the electronic structure as these can qualitatively change the properties of a molecular system.\cite{pyykko-arpc-63-45-2012,schwerdtfeger-rel-p1,schwerdtfeger-rel-p2,dyall-faegri-2007,reiher-wolf-2009} The bonding analysis in such systems is especially demanding if the spin-orbit coupling (SOC) is non-negligible.\cite{dognon-ccr-344-150-2017} There is currently intense activity in the development and practical use of new analysis tools and methods that can reliably characterize chemical bonding irrespective of the strength of SOC, and as we hope to show, TDA is a promising approach.
 
In order to illustrate and further extend the idea of using TDA for interatomic interactions, we have chosen a molecular system studied recently in ref.\cite{anderson-cej-25-2538-2019},
Au$_4$-S-C$_6$H$_4$-S'-Au'$_4$ (\ref{fig_mol}), which is interesting in the context of gold-sulfur\cite{hakkinen-natcom-4-443-2012,pensa-acr-45-1183-2012,xue-natcom-5-4348-2014,rodriguez-cpl-660-287-2016,rodriguez-cpl-732-136625-2019} and gold-hydrogen\cite{rigoulet-pnas-116-46-2018,straka-acie-58-2011-2019,schmidbaur-acie-58-5806-2019,schmidbaur-csr-43-345-2014,verma-organometallics-38-2591-2019} interactions due to their ubiquity in metalorganic complexes and on interface - a proof of many faces of the fascinating chemistry of gold.\cite{bourissou-natchem-11-199-2019,pyykko-acie-43-4412-2004,pyykko-ica-358-4113-2005,pyykko-cr-37-1967-2008} A remarkable finding of the study of Anderson and co-workers\cite{anderson-cej-25-2538-2019} is that it characterized the emergence of NCIs between gold and hydrogen centers, using the Quantum Theory of Atoms-In-Molecules (QTAIM).\cite{bader-qtaim-book-1994,bader-cr-91-893-1991}

QTAIM is a very popular analysis tool,\cite{popelier-chembond-2014} which describes the stationary points of the ED gradient and interprets these points as chemical objects - maxima correspond to nuclei (although some molecules exhibit non-nuclear attractors\cite{gatti-tca-72-433-1987}), while saddle points are identified as \emph{bond critical points (BCP)}, \emph{ring critical points (RCP)} or \emph{cage critical points (CCP)}. ED gradient paths emanated from a saddle point, terminating at two maxima and traced along maximum ED, originally named \emph{bond paths},\cite{bader-jpca-102-7314-1998} are interpreted as interaction lines between atoms and the areas of space delimited by zero-flux surfaces of ED gradient define \emph{atomic basins}. However, recent ambiguous results raised a discussion on the limitations of this method,\cite{cerpa-cej-14-10232-2008,ponec-inorgchem-48-11024-2009,mebs-inorgchem-50-90-2011,lane-jctc-9-3263-2013,foroutan-nejad-cej-20-10140-2014,wick-jmm-24-142-2018,jablonski-chemopen-8-497-2019,politzer-sc-2019,brown-structchem-2019,wilson-cjc-2019} an important point in this debate being precisely the failure of QTAIM to study NCIs. 

While it can be particularly appealing to use QTAIM in this context, especially due to its availability in many standard quantum chemistry codes and supposedly simple interpretation of topological features of ED, it is known that the latter is not a sensitive descriptor of subtle interatomic interactions, due to its exponentially decaying behavior away from the nuclei.
This observation has led to the development of other descriptors, better suited for studying non-covalent interactions, such as the NCI index\cite{johnson-jacs-132-6498-2010,contreras-garcia-jctc-7-625-2011,otero-de-la-rosa-pccp-14-12165-2012,narth-topochem-491-2016} based on the reduced density gradient (RDG), $s(\vec{r})$:
\begin{equation}
s(\vec{r}) = \frac{1}{2(3\pi^2)^{1/3}}\frac{|\nabla\rho(\vec{r})|}{\rho(\vec{r})^{4/3}}.
\label{eq:redgrad}
\end{equation}
At critical points of ED where $\nabla\rho(\vec{r}) = 0$, RDG reaches its minimum value, $s(\vec{r}) = 0$, therefore tracing the minima of $s(\vec{r})$ can be used to locate the interaction sites in molecular systems. The minima of $s(\vec{r})$ may also appear in areas not associated with QTAIM critical points - referred to as \emph{"non-AIM critical points"} or \emph{"interaction critical points (ICPs)"}\cite{contreras-garcia-tca-135-242-2016,boto-tca-136-139-2017} - in which case they indicate very weak interactions.
 
In practical applications, both ED and RDG are approximate and their quality depends on the quantum mechanical (QM) model adopted in calculations. The approximations introduced with this model,\cite{hehre-motheory-book-1986,pople-rmp-71-1267-1999,saue-cpc-12-3077-2011} involving the choice of the Hamiltonian (determining whether and which relativistic effects are included), the method (responsible for the description of electron correlation) and the basis set (establishing the quality of one-electron functions), as well as other numerical parameters (such as the numerical thresholds, grids), can be assessed in a systematic way, yet there are relatively few studies illustrating how all of them affect the aforementioned fields and their topological features.\cite{eickerling-jctc-3-2182-2007,pilme-jctc-10-4830-2014,bucinsky-jpca-120-6650-2016,anderson-jcc-38-81-2017}

Inaccuracies of QTAIM results also stem from theoretical and semantic reasons. The former is related to the fact that QTAIM uses concepts from the graph theory, hence inherits its limitations, in particular it can give the "static" or "binary" information about topological features of ED, but not on their evolution, stability or importance (contrary to TDA). The latter reason is a source of continuous debate, related to the misinterpretation of QTAIM concepts (an example of which is a confusion between \emph{"bond paths"} and \emph{"chemical bonds"}\cite{bader-jpca-113-10391-2009,shahbazian-cej-24-5401-2018,politzer-sc-2019,deLange-pccp-21-20988-2019}), and to the overuse of the adjective \emph{"topological"} in chemistry which should not be confused with mathematical topology.\cite{ayers-ctc-1053-2-2015,mezey-aipcp-1906-020001-2017}

The TDA approach to the interatomic interactions is more general. TDA provides information on the topological features of a scalar field, including their properties, configuration, evolution and significance, while staying oblivious to what the scalar field represents. This detachment of a tool and an interpretation of its output is very helpful to gain a new perspective on the research problem or to redefine existing concepts in a more robust way. In particular, TDA applied to ED, RDG and a signed electron density ($\tilde{\rho}(\vec{r}) = sign(\lambda_2(\vec{r}))\rho(\vec{r})$ with $\lambda_2$ being the second eigenvalue of the ED Hessian)\cite{contreras-garcia-jctc-7-625-2011} enabled to automatically extract and classify interatomic interactions merely based on the \emph{topological persistence} (\emph{vide infra}) of critical point pairs of these fields.\cite{gunther-ieeetvcg-20-2476-2014}

\section{Computational details}

\begin{figure}[h!]
 \centering
 \includegraphics[width=.6\linewidth]{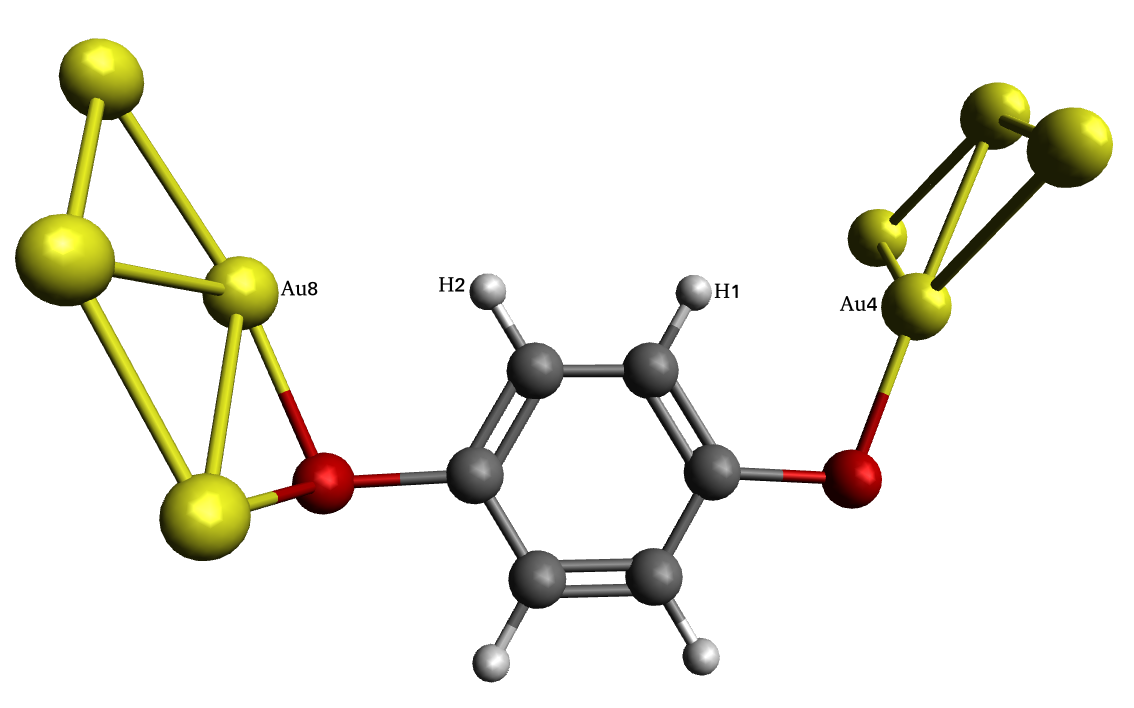}
 \includegraphics[width=.3\linewidth]{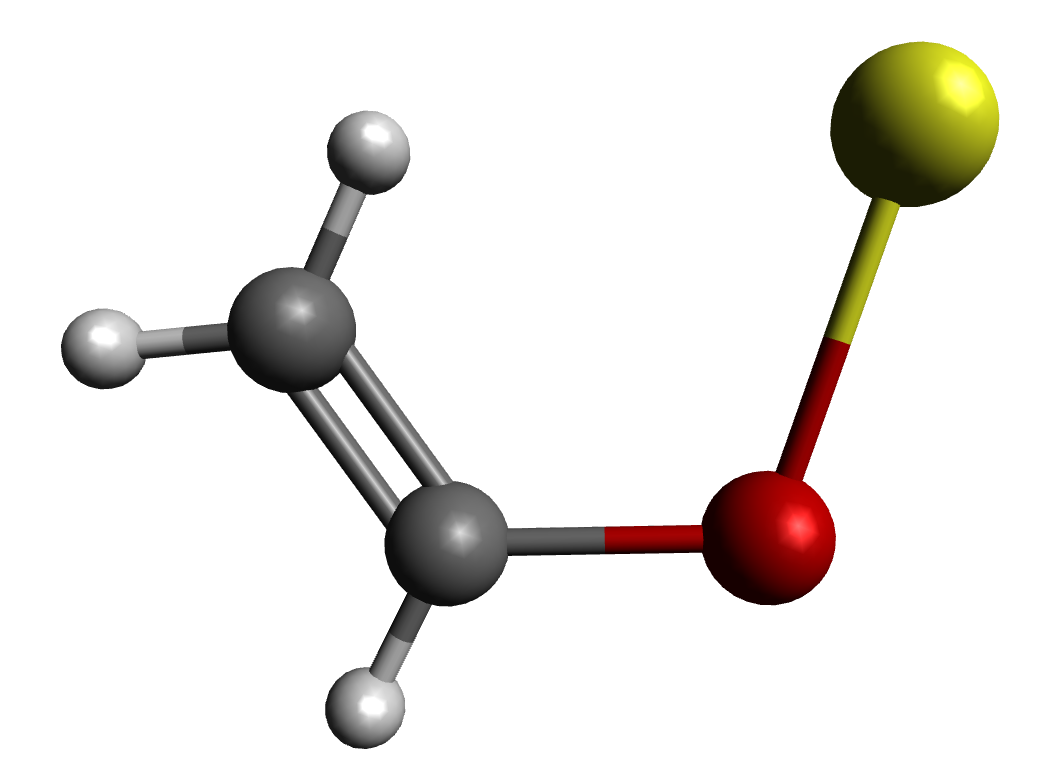}
 \caption{Molecular systems under study: Au$_4$-S-C$_6$H$_4$-S'-Au'$_4$ (left) and Au-S-CH-CH$_2$ (right).
 The numbering of selected Au and H centers is shown and is the same as in ref.\cite{anderson-cej-25-2538-2019}.
 Color coding for atoms: Au - yellow, S - red, C - dark gray, H - light gray.
 Additional information is available in ESI.}
 \label{fig_mol}
\end{figure}

Following ref.\cite{anderson-cej-25-2538-2019}, the geometry of the Au$_4$-S-C$_6$H$_4$-S'-Au'$_4$ molecule was optimized in the ADF software\cite{ADF2001,ADF2017authors,guerra-tca-99-391-1998} using three different models:
(i) scalar-relativistic Zeroth-Order Regular Approximation (sr-ZORA),\cite{vanlenthe-jcp-105-6505-1996,vanlenthe-jcp-101-9783-1994,vanlenthe-ijqc-57-281-1996,vanlenthe-jcp-110-8943-1999} PBE functional\cite{perdew-prl-77-3865-1996} and TZP basis set\cite{vanlenthe-jcc-24-1142-2003} at the singlet spin-restricted level (denoted throughout the text as \textbf{g1}), (ii) sr-ZORA, PBE, TZP at the triplet spin-unrestricted level (\textbf{g2}) and (iii) sr-ZORA, PBE, QZ4P\cite{vanlenthe-jcc-24-1142-2003} at the triplet spin-unrestricted level with tighter SCF convergence criteria (\textbf{g3}). The study of a triplet state of this molecule, motivated by its proximity to the singlet ground state, was conducted to explore how the spin state affects the ED topology.

We considered simpler model systems of that molecule to investigate the importance of a local environment to the topological features of interest, Au-S-CH-CH$_2$ (\ref{fig_mol}). Its geometry was obtained (without reoptimization) by removing atoms from the \textbf{g1} model of Au$_4$-S-C$_6$H$_4$-S'-Au'$_4$ molecule, so that it contains only one gold atom linked by sulfur to the ethene group that replaced the benzene ring. As the positions of all nuclei in Au-S-CH-CH$_2$ were preserved as in the original molecule and due to the lack of symmetry in Au$_4$-S-C$_6$H$_4$-S'-Au'$_4$, this simplification resulted in two geometries of the small model (atom numbers in parenthesis, after ref.\cite{anderson-cej-25-2538-2019}) - Au(4)-S(1)-C(1)H-C(2)H(1)H ("a") and Au(8)-S(2)-C(4)H-C(3)H(2)H ("b"), hence the nomenclature \textbf{m-a-g1} and \textbf{m-b-g1}, respectively.

\begin{figure*}
 \centering
 \includegraphics[width=0.95\linewidth]{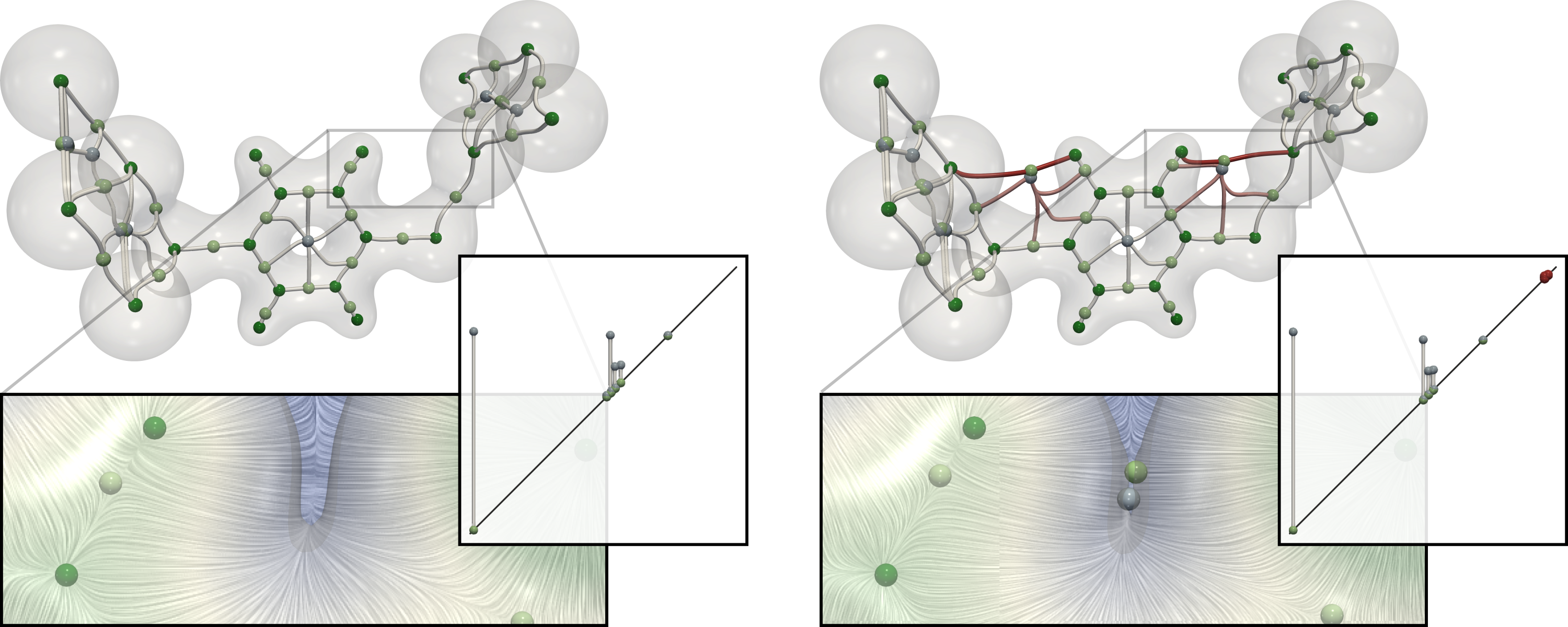}
 \caption{Comparison of the topological structure (one-dimensional
separatrices and persistence diagram of the one-dimensional persistent homology
group) for the logarithm of the electron density, $log(\rho)$,\cite{whylog}
of
the Au complex without (left) and with (right) relativistic effects.
Isosurfaces of
$log(\rho)$
(for the same isovalues) are shown in
transparent gray to illustrate the geometry of the molecule.
Top: the critical points of
$log(\rho)$
are shown with colored spheres
(green: maxima, light green: $2$-saddles, light blue: $1$-saddles) and integral
lines connecting pairs of critical points are shown with white curves. With
relativistic effects (right), an additional $1$-saddle/$2$-saddle pair emerges,
indicating the possible existence of a hydrogen bond (shown in dark red).
Bottom: zoom on the corresponding regions (background: line integral
convolution of $\nabla log(\rho)$,
colored with $log(\rho)$ blue to green). With
relativistic effects (right), the isosurface changes its genus
($\beta_1$\cite{betti-def}) in
the vicinity of the extra saddle-saddle pair (tiny handle in the surface
around the pair of light green and light blue spheres). Without relativistic
effects (left), the isosurface (for the same isovalue) does not exibit any
handle and $log(\rho)$ admits no saddle-saddle pair in this region of space.
The
persistence diagram \cite{edelsbrunner09} (foreground
insets) of the one-dimensional persistent
homology group of the sub-level sets of $log(\rho)$  displays each
saddle-saddle pair $(s_1, s_2)$ in the 2D plane at coordinates
$\Big(log\big(\rho(s_1)\big), log\big(\rho(s_2)\big)\Big)$. The height to the
diagonal ($log\big(\rho(s_2)\big) -
log\big(\rho(s_1)\big)$, white bar) denotes an importance measure called in
Topological
Data Analysis the \emph{Persistence} of the critical point pair
(illustrating the lifespan in the data of the corresponding topological
feature). On the right diagram, the extra saddle-saddle pair is displayed in
dark red (rightmost spheres). Its extremely low persistence indicates a very
weak salience of the corresponding topological feature, questioning its
interpretation as a hydrogen bond suggested in Anderson et al.\cite{anderson-cej-25-2538-2019}}
 \label{fig_comparison}
\end{figure*}

For all studied molecules, ED has been evaluated in single-point calculations with the ADF software. These employed the non-relativistic (NR) Hamiltonian as well as scalar and spin-orbit ZORA Hamiltonians (sr-ZORA and so-ZORA, respectively), DFT with different exchange-correlation (XC) functionals (PBE, B3LYP,\cite{stephens-jpc-98-11623-1994} SAOP,\cite{schipper-jcp-112-1344-2000,gritsenko-cpl-302-199-1999} M06-2X\cite{zhao-tca-120-215-2008,zhao-jcp-125-194101-2006}) and three basis sets (DZ,\cite{vanlenthe-jcc-24-1142-2003} TZP and QZ4P). Additionally, for the small Au model we performed calculations in the locally-modified version of the DIRAC software\cite{DIRAC18} with the L\'evy-Leblond (LL),\cite{levy-leblond-cmp-6-286-1967} spinfree Dirac-Coulomb (SFDC)\cite{dyall-jcp-100-2118-1994} and Dirac-Coulomb (DC)\cite{visscher-tca-98-68-1997} Hamiltonians, two methods involving different levels of electron correlation (Hartree-Fock (HF) and DFT with the PBE functional) and a basis set of triple-zeta quality.\cite{dyall-tca-129-603-2011,dyall-tca-131-1217-2012,dirac-basis,woon-jcp-98-1358-1993}

The TDA of the logarithms of ED and RDG ($log(\rho)$ and $log(s)$, respectively\cite{whylog}) was performed with the TTK software package.\cite{TTK}
All details can be found in the electronic supplementary information (ESI).

\section{Results and discussion}
Even though the optimized geometries of the Au complex are slightly different
than in ref.\cite{anderson-cej-25-2538-2019}, our analysis confirms the main 
conclusions of the cited work: additional pairs of critical points of ED (1-saddles and 2-saddles, interpreted as \emph{"RCPs"} and 
\emph{"BCPs"}, respectively) between Au and H nuclei appear for certain combinations of Hamiltonians, 
methods, basis sets, molecular geometries and spin states, but they never appear 
in the non-relativistic context. 

The comparison of topological features of 
relativistic (so-ZORA) and non-relativistic EDs of \textbf{g1} molecule obtained with the PBE/QZ4P QM 
model is illustrated in 
\ref{fig_comparison}, where, in particular, the topological skeleton of non-relativistic ED 
(without these additional saddle-saddle pairs) and relativistic ED (with these pairs) 
are shown. The results from TDA of these densities are described in detail in the caption to \ref{fig_comparison}.

The TDA pattern of critical points and 1-separatrices of ED is 
principally the same as the one obtained from QTAIM 
analysis.\cite{gunther-ieeetvcg-20-2476-2014} However, as we shall show, TDA 
of ED calculated with various QM models paints a more nuanced picture.

\begin{figure}
 \centering
 \includegraphics[width=.95\linewidth]{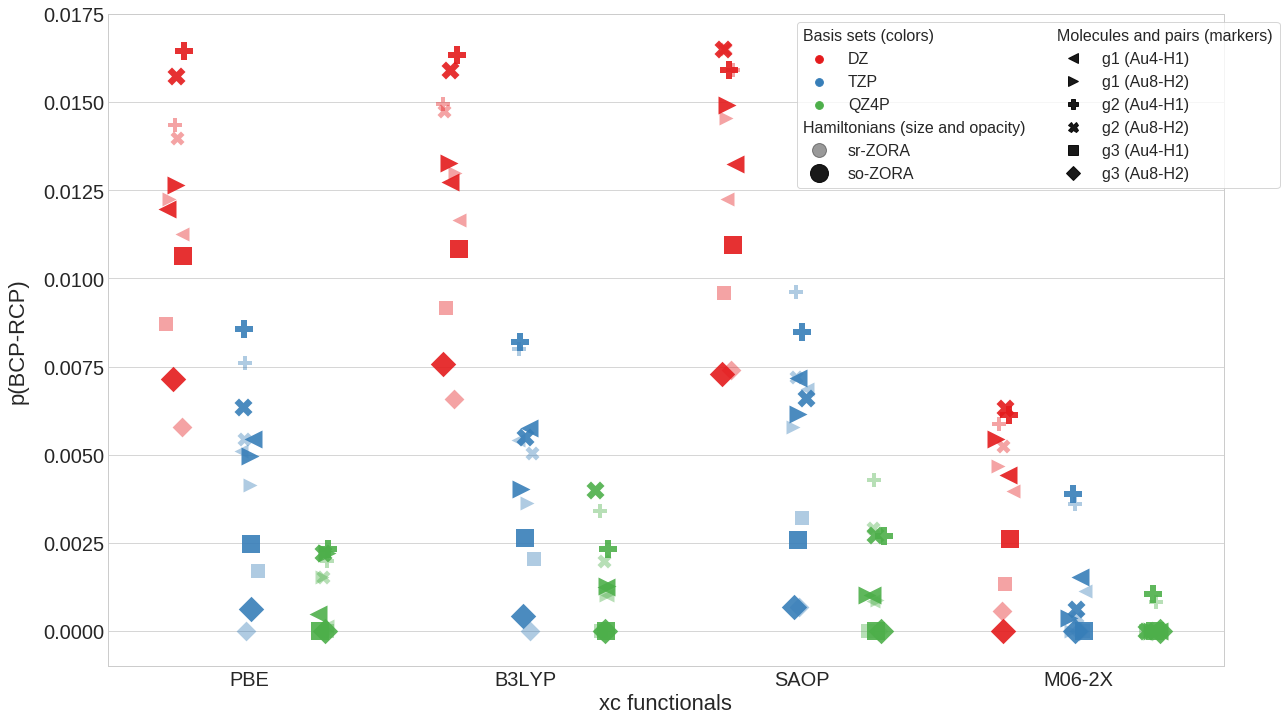}
 \vfill
 \includegraphics[width=.95\linewidth]{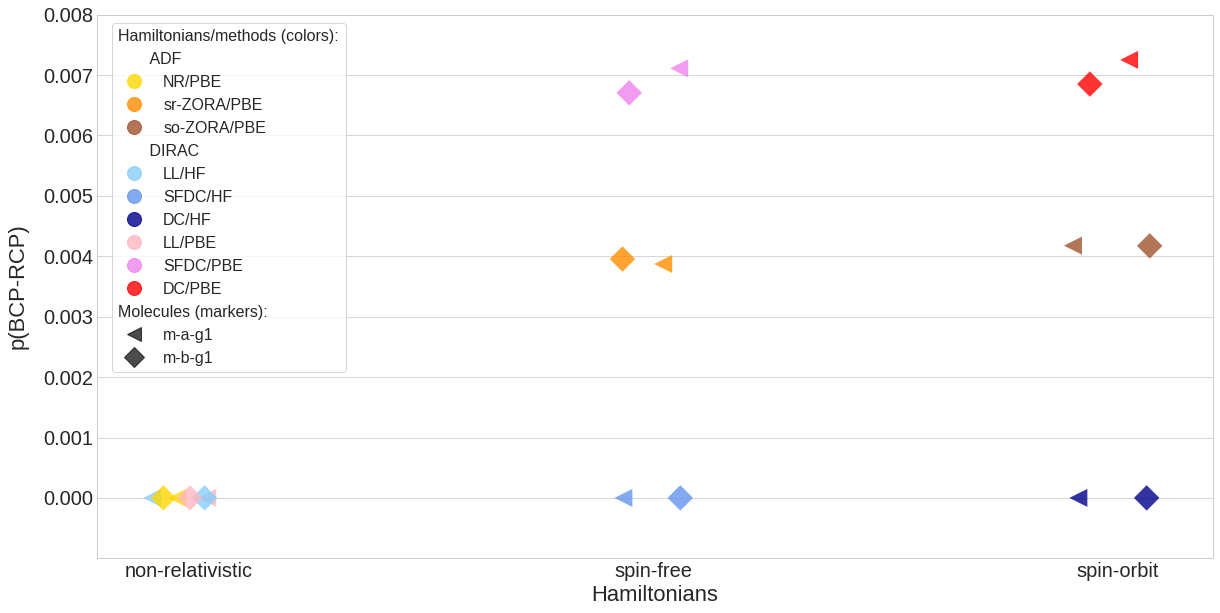}
 \caption{Persistence values of the extra saddle-saddle pairs of $log(\rho)$ between H1 and Au4 and between H2 and Au8 atoms in Au$_4$-S-C$_6$H$_4$-S'-Au'$_4$ molecule (top) and between Au and H atoms in
 Au-S-CH-CH$_2$ molecule (bottom) calculated with different QM models.
 Top: the plot demonstrates that the persistence values of interesting pairs can be clustered by the basis set used in ED calculations, 
 indicating that overall it is the basis set quality that affects the ED topology to a largest extent, 
 while the choice of the XC functional (within tested ones), 
 the inclusion of the spin-orbit coupling (at the ZORA level) or 
 the chemical environment of the extra
 saddle-saddle pair do not have that significant effect.
 Bottom: the plot exposes the role of electron correlation evaluated at the DFT/PBE level in the DIRAC software 
 and shows the differences in results obtained with ADF and DIRAC codes.
 The persistence values of absent saddle-saddle pairs are set to zero (\ref{eq:pers}).
 Numerical values are available in ESI.
 }
 \label{fig_persistence_allmols}
\end{figure}

Before continuing on that thought, we shall clarify that the presence of 
1-saddles accompanying 2-saddles between Au and H atoms is 
a result of Morse inequalities\cite{milnor63}
\footnote{Equivalently it can be deduced from the Poincar\'e-Hopf theorem, which applies to generic vector fields.}
and is in accordance with chemical intuition due to 
the ring-shaped geometry of the Au-S-C-C-H chains.
In Persistent Homology,
the critical points of a scalar function 
are paired in a non-ambiguous way (called the Elder rule\cite{edelsbrunner09}) 
and in this particular case, the additional 1-saddle and 2-saddle points, if present, form a pair on 
each side of the aromatic ring. The \emph{lifespan} of this pair is measured by 
\emph{persistence}, calculated as a difference of scalar field values in
corresponding critical points. 
Accordingly, we define the persistence for the additional \emph{"BCP-RCP"} pair 
as the difference of the (logarithm of) densities in the position of relativistic \emph{"BCPs"} and \emph{"RCPs"} as:
\footnote{
The 
mathematical background of persistence can be found 
elsewhere.\cite{edelsbrunner09}
} 
\begin{align}
 p_{BCP-RCP} &= \nonumber \\ 
 &  \begin{cases} 
   | log(\rho_{BCP}) - log(\rho_{RCP})|    & \exists~\text{\emph{BCP-RCP}} \\
   0                                       & \nexists~\text{\emph{BCP-RCP}}
  \end{cases}
  \label{eq:pers}
\end{align}
and use this definition to calculate the persistence values of 
extra saddle-saddle pairs of $log(\rho)$ determined with various QM models. 
In all cases these values are very small, constituting typically less than 0.1\% of the full persistence range (represented by red bars on the persistence diagram plotted for relativistic ED in \ref{fig_comparison}).
Furthermore, they decrease with an increasing accuracy of the QM models, notably in molecules 
\textbf{g1}, \textbf{g2} and \textbf{g3}, these persistence values 
systematically reduce along the DZ - TZP - QZ4P basis sets order for all XC 
functionals (\ref{fig_persistence_allmols}), a trend which is is observed for both spin states and all geometries.
The most remarkable change is seen for the combination of  QZ4P basis and the meta-GGA M06-2X functional, for which these  
 saddle-saddle pairs disappear (as signified by the zero-persistence values on 
the plots for all but the Au4-H1 pair in the \textbf{g2} system. 
 This functional is also the one for which  persistence values are generally lower. Taking at face value the claim of its superior performance for describing non-covalent interactions compared to the other functionals used in this work,\cite{zhao-tca-120-215-2008,bircher-jctc-15-557-2019} our findings that this functional gives zero-persistence values would therefore imply the absence of these topological features.

At this point it is also interesting to compare results for \textbf{g2} and \textbf{g3}, since both represent the triplet state of the Au$_4$-S-C$_6$H$_4$-S'-Au'$_4$ molecule, only with a larger basis set and tighter SCF criteria used in the geometry optimization and ED calculations of the latter.
It turns out that these improvements in the calculations have a tremendous effect on the persistence values of extra saddle-saddle pairs on both sides of the aromatic ring, namely these pairs are absent in \textbf{g3} in the most accurate setups (QZ4P basis set) for all selected XC functionals.
This confirms the conclusion from ref.\cite{anderson-cej-25-2538-2019} that the topology of ED significantly changes depending on how close the structure is to a minimum of the Born--Oppenheimer (BO) energy surface, but also underscores that this pair is not as significant as the conventional QTAIM analysis would make it to be.

Similar analysis of \textbf{m-a-g1} and \textbf{m-b-g1} (\ref{fig_persistence_allmols}), 
this time focusing on the comparison of different Hamiltonians and methods,
reveals the strong influence of the electron correlation on the persistence values of the additional saddle-saddle pair, seemingly counteracting the relativistic effects. 
This appears from the DIRAC calculations involving the fully-relativistic 4-component DC Hamiltonian with HF and DFT/PBE methods: the ED calculated at HF level does not exhibit the extra saddle-saddle pair, while this pair is present in the DFT calculations and its persistence is (slightly) larger than when an approximate relativistic Hamiltonian (SFDC) is used.
This observation, together with the negligible role of the spin-orbit coupling on the persistence values (as evaluated on scalar-relativistic geometries),
suggests further studies on \textbf{m-a-g1} and \textbf{m-b-g1} with wavefunction-based methods at the scalar-relativistic level of theory.
Apart from a change in electronic structure method, it may well be that changes in equilibrium structures take place with the consideration of spin-orbit coupling in the geometry optimization procedure, as evidenced by a recent study on SOC-triggered bond contraction in gold clusters.\cite{flores-jp-2016} We have nevertheless opted in this work to focus on selected geometries such as those previously considered by Anderson et al.\cite{anderson-cej-25-2538-2019} This was motivated, first, by the comparison of the results for these model systems with the original molecular system, which indicates that the chemical surrounding of interacting Au and H atoms does not necessarily affect the topology of ED in these Au-H interatomic areas. And, second, that we considered that this would divert attention to what we believe is the central point of this work, which is to underscore the differences and similarities between TDA and QTAIM. We note, however, that further studies on these gold systems should indeed consider the effect of structural chances brought about by SOC.

In TDA the consensus is that the persistence of a topological feature is associated with 
its \emph{importance} (in particular in the presence of additive noise), 
therefore it is frequently used in the topological 
simplification of complex scalar fields. Features manifested by so small 
persistence values would typically be deleted, which raises the question of 
whether the extra \emph{"BCPs"} and \emph{"RCPs"} of ED are an important feature of this scalar 
field.
The answer to 
this
question relies on the domain knowledge to provide interpretation of such feature. In the light of late research which cast \emph{"serious doubts on the interpretation of BCPs in QTAIM analyses as indicating bonding interactions, however weak"},\cite{wick-jmm-24-142-2018} confronted with the recently proposed translation of these additional saddle-saddle pairs as indicators of \emph{"favorable interaction"},\cite{anderson-cej-25-2538-2019} we are inclined to answer "no". 

There are however more arguments standing for this answer, which do not involve the chemical interpretation of these additional critical points.
For instance, the line integral convolution plots of the ED gradient demonstrated in \ref{fig_comparison} have a very similar pattern in relativistic and non-relativistic contexts, what is understandable considering that there need to be a border separating the atomic basins of Au and H (regardless of the QM model). In addition, as also exposed in \ref{fig_comparison}, the Euclidean distance between the extra "RCP" and the extra "BCP" in relativistic case is so small, that their presence can be considered as a negligible (\emph{unimportant}) perturbation to the topology of ED in this interatomic area.

\begin{figure*}
 \centering
 \includegraphics[width=\linewidth]{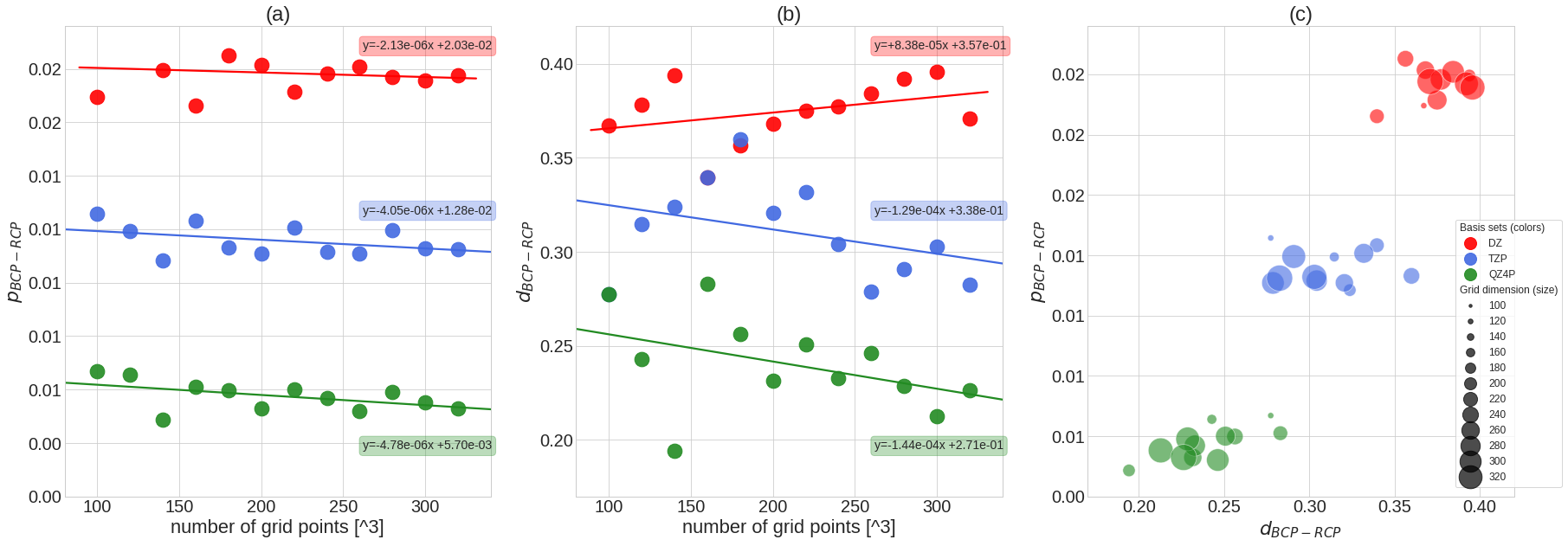}
 \caption{Au-S-CH-CH$_2$ molecule (\textbf{m-a-g1} model): 
 (a) the dependence of the persistence values of the extra saddle-saddle pair of $log(\rho)$ 
 between H and Au atoms ($p_{BCP-RCP}$) on the grid size, 
 (b) the dependence of the Euclidean distance between these saddle points ($d_{BCP-RCP}$) 
 on the grid size and 
 (c) the correlation of $p_{BCP-RCP}$ and $d_{BCP-RCP}$ values, 
 with the dependence on the basis set and the grid size marked on the plot.
 Plots (a) and (b): the linear regression models were used to
estimate a simple relationship between variables. 
The actual persistence values
of the additional saddle-saddle pairs of $\rho$ decay exponentially with the
grid size (note the $log(rho)$ scalar field under analysis),
which suggests that the topological features under study may be artificial.
The comparison of slopes of the fitted lines in DZ-TZP-QZ4P cases
highlights the faster
convergence of $p_{BCP-RCP}$ and $d_{BCP-RCP}$ with the grid size in the most
accurate QM setup (QZ4P).
Plot (c): clear data clustering with respect to basis
set further emphasizes the importance of the basis set in QM-based ED
topology analysis.
Numerical values are available in ESI.
Color coding common for all plots: DZ - red, TZP - blue, QZ4P - green markers.}
 \label{fig_m4_grid}
\end{figure*}

In order to test the robustness of the presented analysis with respect to grid sampling, we have analyzed the topology and geometry of ED of \textbf{m-a-g1} molecule calculated with a selected QM model (so-ZORA Hamiltonian, PBE functional and three basis sets, DZ, TZP and QZ4P) on a series of regular grids of size ranging from 100$^3$ to 320$^3$ points. The results of this analysis are demonstrated in \ref{fig_m4_grid} and they confirm prior observations: the extra saddle-saddle pairs (if present) can indeed be considered as unimportant topological features of ED, if not artificial due to the expected exponential decay of the persistence values of the extra pairs of the $\rho$ scalar field with the grid size (linear dependence of the persistence values of the corresponding pairs exhibited by the $log(\rho)$ scalar field).

What clearly discerns the relativistic ED exhibiting \emph{"BCP-RCP"} pairs and nonrelativistic EDs without these pairs (for the same QM setup) is the fact that in the former case $\rho^{R}(r^{R}_{BCP}) > \rho^{R}(r^{R}_{RCP})$, what causes the change of genus of the ED surface when sweeping over all values of ED, manifested by the appearance of additional critical points, as demonstrated in \ref{fig_comparison}. 
In the latter case, the value of ED probed in the location of 
\emph{"relativistic BCP"}, $r^{R}_{BCP}$, is lower than the value of ED in the 
position of the \emph{"relativistic RCP"}, $r^{R}_{RCP}$, hence 
$\rho^{NR}(r^{R}_{BCP}) < \rho^{NR}(r^{R}_{RCP})$. Consequently, the surface 
does not change its genus and the critical points do not appear. Yet, the 
difference of density values in these two locations ($r^{R}_{RCP}$ and 
$r^{R}_{BCP}$) is very small (Table 13 in ESI), which calls for a more accurate 
analysis involving properties which are based on the higher-order derivatives of 
the ED, such as the ED gradient or the ED Hessian.\cite{eickerling-jctc-3-2182-2007}
Other QM approximations, such as the ones determining the description of electron correlation or the quality of one-electron functions do not have such a strong influence on ED that could effectively change the total order of density values in these two locations, however as the final picture is a result of interplay of many effects (e.g. counteracting relativistic corrections and electron correlation), none of them should be neglected. Notably, relativistic formalism should be used if the interacting sites involve heavy elements, as already discussed elsewhere.\cite{bucinsky-jpca-120-6650-2016,eickerling-jctc-3-2182-2007,anderson-cej-25-2538-2019}

\begin{figure*}
 \centering
 \includegraphics[width=.45\linewidth]{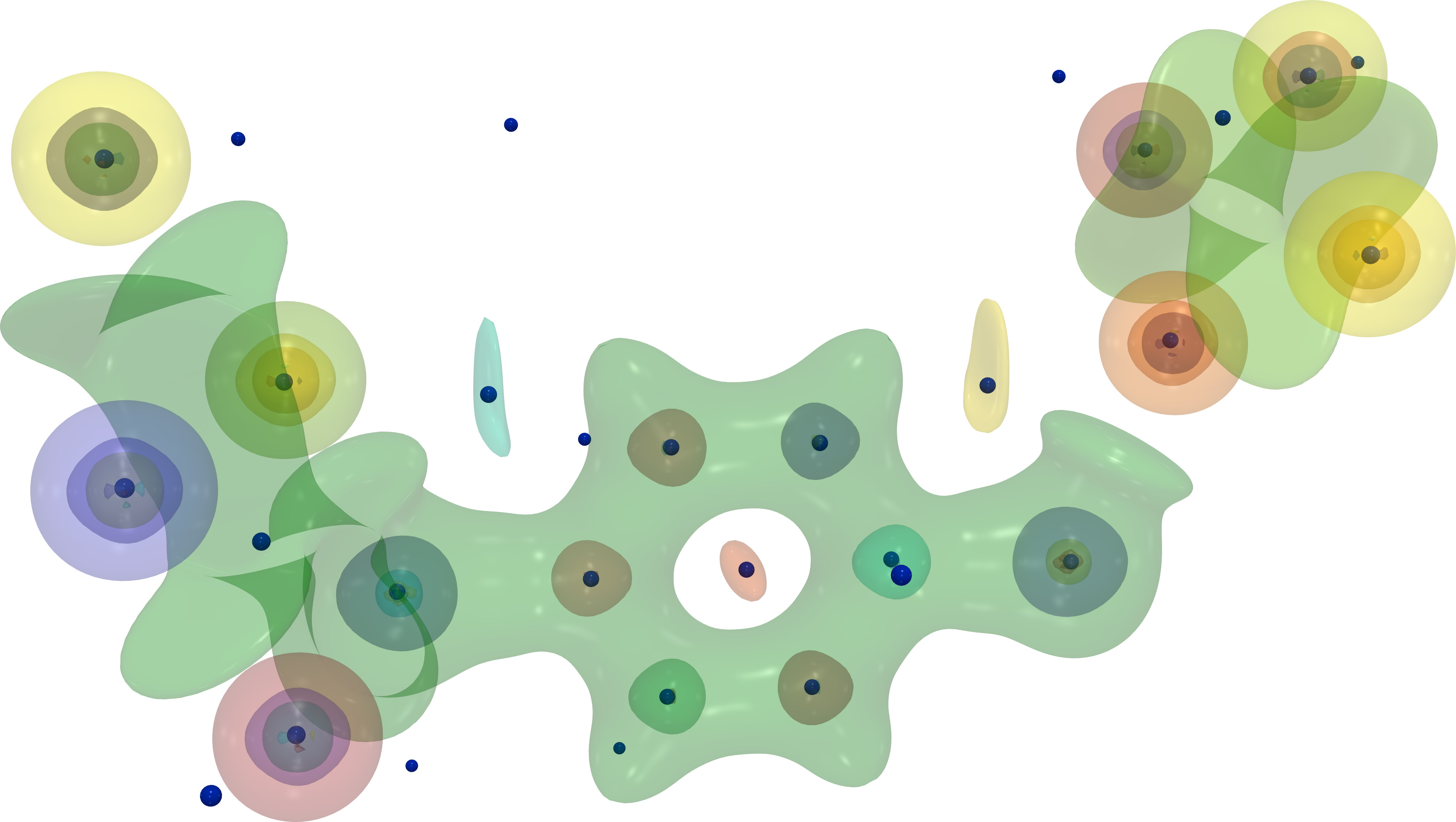}
 \hfill
 \includegraphics[width=.45\linewidth]{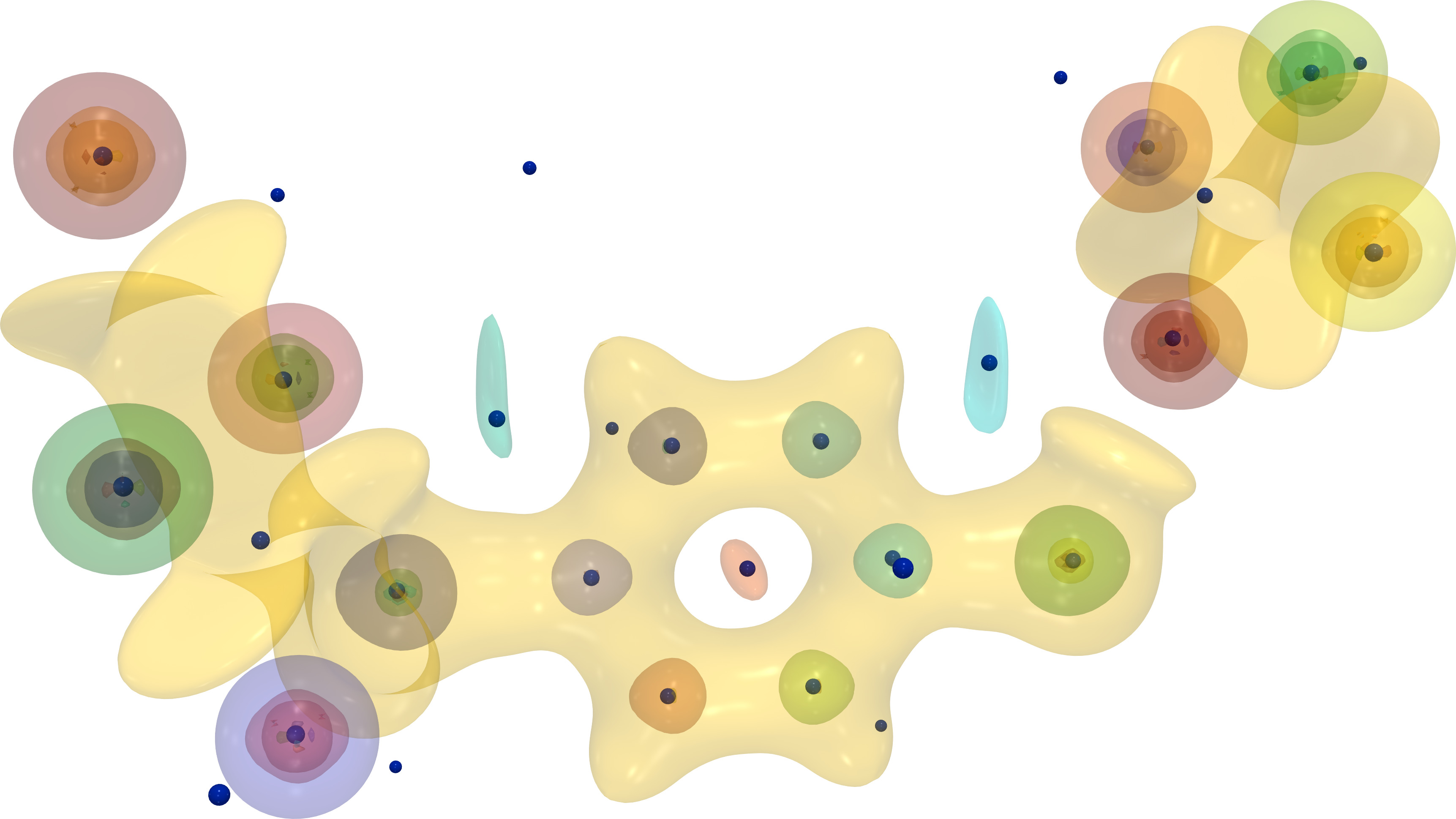}
 \caption{Comparison of data segmentations based on isosurfaces of reduced
density gradient (same isovalue) without (left) and with (right) relativistic
effects for the Au complex. Each connected component of isosurface is displayed
with a unique, randomly chosen, color.
The number of connected components is equal in the two cases, and their
geometry is highly similar. In particular, the regions exhibiting extra
saddle-saddle pairs of $log(\rho)$
(\ref{fig_comparison}) contain isosurface
components with very similar geometry in both cases.}
 \label{fig_comparison_reducedGradient}
\end{figure*}

The use of accurate QM setup is essential to assert a good quality of calculated ED, however it does not change the fact that one should be cautious when drawing conclusions on bonding interactions based on the QTAIM output only.
We have therefore continued with the analysis of the RDG in studied systems. As illustrated in \ref{fig_comparison_reducedGradient}, there is a striking similarity between the relativistic and non-relativistic RDGs, in geometrical (the shape of the RDG isosurfaces) and topological (the same number of connected components ($\beta_0$\cite{betti-def}) and the presence and location of minima) senses. In particular, the interaction sites between Au and H atoms are present in both contexts. This is an additional argument for the resemblance of Au-H non-covalent interactions with or without relativistic effects included.

\section{Conclusions}
In conclusion, we have used TDA to reinterpret the QTAIM results for gold complexes reported recently,\cite{anderson-cej-25-2538-2019} in which relativistic effects are considered to be behind the emergence of new 1-saddle - 2-saddle point pairs between non-covalently bonded Au and H atoms. 

The choice of TDA is motivated by its favorable performance (by using combinatorial algorithms instead of numerical grid searches) and mathematical robustness, allowing for an explicit translation between topological features and their chemical interpretation. TDA offers the possibility of visualizing the "topological skeleton" of a scalar field, including for the EDs the same features as QTAIM (critical points and 1-separatrices between 2-saddles and maxima that correspond to "bond paths") as well as others that do not have their QTAIM correspondence (1-separatrices between 1-saddles and 2-saddles, 2-separatrices, manifolds of the Morse-Smale complex and outputs from other TDA data abstractions).

Our results demonstrate that the relationship between the topology of ED and the existence of non-covalent interactions is more complex than what can be deduced from a QTAIM picture based on the presence or absence of \emph{"BCPs"} and \emph{"bond paths"} between atoms. While TDA confirms that the additional saddle-saddle pairs between Au and H atoms are indeed the features of the relativistic ED which are absent for a non-relativistic ED, these feature appear not to be meaningful, due to their low persistence values and, more importantly, to the fact that these values approach zero with increased sophistication of the calculations (larger basis set and denser grids). 

Furthermore, our observations are another example of the by now established fact that the ED is not the best descriptor of such weak interactions, for which other quantities such as the RDG are better suited.\cite{lane-jctc-9-3263-2013,contreras-garcia-tca-135-242-2016,boto-tca-136-139-2017,peccati-ctc-1159-23-2019} Our geometrical and topological analysis of the RDG, however, does not show qualitative differences between the non-relativistic and relativistic calculations, a fact which further calls into question the importance of relativistic effects to non-covalent interactions in such systems.

Taken together, these conclusions point to the need for a deeper analysis of the influence of the relativistic effects on the non-covalent interactions. To this end, we have shown here that much smaller model systems retain the same topological features as the original molecules, paving the way for more accurate calculations, such as those based on correlated wavefunction methods.

Finally, we suggest that TDA, and notably the concept of persistence--which can be considered as a process of transforming discrete invariant into a continuous one\cite{chacholski-topoinvis}--should be considered in the development of real-space analysis techniques of the chemically-relevant scalar fields.

\section*{Acknowledgements}

We acknowledge support from the Polish National Science Centre (NCN) (grant 
number 2016/23/D/ST4/03217), the Labex CaPPA (contract ANR-11-LABX-0005-01) and 
I-SITE ULNE project OVERSEE (contract ANR-16-IDEX-0004), CPER CLIMIBIO (European 
Regional Development Fund, Hauts de France council, French  Ministry of Higher 
Education and Research), the French national supercomputing facilities (grant 
DARI A0050801859), and the European Commission grant H2020-FETHPC-2017 
``VESTEC'' (ref. 800904).
We would like to thank Dr Val\'erie Vallet for her comments on the draft of this paper.

\section*{Conflict of interest}
There are no conflicts to declare.

\section*{Supporting Information Available:} 

The supporting information is available on zenodo, \href{https://zenodo.org/record/3358788#.XURPL9_nhhG}{DOI:10.5281/zenodo.3358788}.

\bibliography{manuscript}

\begin{thebibliography}{100}
\bibitem{whylog}
The logarithm scale was employed in the TDA of ED since the full persistence
  range calculated for the ED is very large, what makes the visualization of
  bars on the persistence diagram corresponding to the extra saddle-saddle
  pairs in its topology (exhibiting very low persistence values) impractical.
  The logarithm scale for the RDG was then used for consistency.
\end{thebibliography}

\end{document}